# Can an amorphous alloy crystallize into a high entropy alloy?

Nikhil Rai, Priyabrata Das, Srikant Gollapudi[*]
School of Minerals, Metallurgical and Materials Engineering
Indian Institute of Technology Bhubaneswar, India 752050

**Abstract:**

On the premise that amorphous-HEA composites could demonstrate high toughness and resistance to embrittlement akin to the phase separating glassy-solid solution composites, we develop a thermodynamics based approach to identify chemical compositions capable of undergoing the amorphous to HEA transformation. We introduce two new parameters called phase selection value (*PSV*) and molar volume dispersity parameter ($\Gamma$). Using this thermodynamic approach seven multi-component compositions were proposed and the general guidelines for identifying such compositions was established. The approach also reveals that BMGs may not be as such amenable to undergo an amorphous to HEA transformation.

We believe Amorphous-High entropy alloy (HEA) composites might provide high toughness in line with the metallic glass-solid solution composites developed by Johnson and co-workers [1,2]. Formation of intermetallics and other complex compounds during thermo mechanical processing in the supercooled region is considered as one of the causes of embrittlement of amorphous alloys [3-6]. Ductile second phases based on HEAs, which are known to bear a remarkable suite of physical properties [7-11], could provide enhanced toughness and hence the important question is, can an amorphous phase transform directly into a HEA leading to development of *in situ* amorphous-HEA composites.

In this letter, we are proposing a thermodynamics based model for identifying compositions capable of an amorphous to HEA transformation. Herein we would like to mention that phase selection between amorphous, *HEA*s and intermetallics has already been investigated and reported in literature [12, 13]. However, unlike the models discussing phase selection between amorphous, *HEA*s and intermetallics, our investigation is focused on checking the capability of an amorphous system to crystallize into a *HEA* as a function of temperature. If *HEA*s can indeed crystallize from the amorphous phase, then the duration and temperature of phase transformation can be optimized to develop amorphous-HEAs composites with desired volume fraction of the *HEA*. Notwithstanding the fact that kinetics of the transformation is a key aspect, our model currently approaches this problem only from a thermodynamics standpoint at this point of time. Guided by this thermodynamics model, we identify seven compositions which we believe are amenable to the amorphous-HEA transformation.

---

[*] Corresponding author
Email: srikantg@iitbbs.ac.in
Ph: 0091 9566288703

In order to determine the possibility of amorphous structure transforming into HEA in preference to intermetallics, it is essential to determine the Gibbs free energy for the three competing phases. For the Gibbs free energy of formation of each of the three phases, it is important to determine the enthalpy of formation and entropy of mixing. The enthalpy of formation for the amorphous phase, solid solution phase and the intermetallic can be determined using the Miedema's model [14, 15]. The enthalpy of formation ($\Delta H_{for}^{ss}$) for a binary solid solution according to the Miedema's model is given by

$$\Delta H_{for}^{ss} = \Delta H^{chem} + \Delta H^{elastic} + \Delta H^{structural} \quad (1)$$

The structural term $\Delta H^{structural}$ is usually very small and ignored during calculations. The other two terms $\Delta H^{chem}$ and $\Delta H^{elastic}$ can be determined using the methodology provided in supplementary section.

For a multi-component system, the enthalpy of formation of a solid solution is an extension of the Miedema's model for a binary system. The enthalpy of formation of solid solution (in this case a *HEA*) for a multi-component system having '*m*' number of elements can be determined using the following equation [16]

$$\Delta H_{for}^{HEA} = 4 \sum_{i,j=1, \; i\neq j}^{m} c_i c_j \Delta H_{mix}^{AB} + \sum_{i,j=1, i\neq j}^{m} \left(\Delta H_{mix}^{elastic}\right)_{i-j} \quad (2)$$

Where $\Delta H_{for}^{HEA}$ is the enthalpy of formation of *HEA*, $\Delta H_{mix}^{elastic}$ is the elastic enthalpy due to atomic size mismatch for any two elements *A* and *B* and *i* and *j* are the *i*th and the *j*th element of the multi-component system. The entropy of the multi-component system is primarily due to configurational entropy and is given by the equation below:

$$\Delta S^{mix} = -R \sum_{i=1}^{m} x_i \ln x_i \quad (3)$$

where $x_i$ is the mole fraction of the *i*th element present in the multi component system. Thus, the Gibbs free energy of high entropy alloy in a multi-component system can be obtained by following equation:

$$\Delta G^{HEA} = \Delta H_{for}^{HEA} - T\Delta S^{mix} \quad (4)$$

Where $\Delta G^{HEA}$ is the free energy of formation of *HEA* and $T$ is the temperature at which the Gibbs free energy is determined. For the amorphous phase, the formation enthalpy ($\Delta H^{am}$) in multi-component system is given by [16]

$$(\Delta H^{am})_{total} = 4\sum_{i,j=1,\ i \neq j}^{m} c_i c_j \Delta H_{mix}^{AB} + \sum_{i=1}^{m} x_i H_i^a \qquad (5)$$

where $H_i^a$ is the enthalpy of amorphization of a pure element and the other terms are as described before. The entropy of the multi-component system in amorphous phase can be obtained by the equation 3. The Gibbs free energy of the amorphous phase can then be determined using equation 3 and equation 5. Between the amorphous and *HEA* phases, the phase having lower value of Gibbs free energy will be the more stable phase and hence their stability vis-à-vis each other can be understood by

$$\Delta G^{HEA} - \Delta G^{am} = (\Delta H_{mix}^{ss})_{total} - (\Delta H^{am})_{total} \qquad (6)$$

where a negative output of equation 6 indicates preference for *HEA* and a positive output indicates preference for amorphous phase. From equation 2 and equation 5, equation 6 can be re-written as:

$$\Delta G^{HEA} - \Delta G^{am} = \sum_{i,j=1, i \neq j}^{m} (\Delta H_{mix}^{elastic})_{i-j} - \sum_{i=1}^{m} X_i H_i^a \qquad (7)$$

For the sake of simplicity, let us call the term $\Delta G^{HEA} - \Delta G^{am}$ in equation 7 as the Phase Selection Value (*PSV*) and this is employed to check for phase selection between HEA and amorphous phase. Herein we would like to clarify two points

a. We are not ignoring the phase selection tendencies of intermetallics and other compounds when checking for amorphous to *HEA* transformation (hereafter referred to as *A-HEA* transformation). We first evaluate the possibility of *A-HEA* transformation using *PSV*. If this transformation is possible for a certain system / composition, then as a next step we check if intermetallics and other compounds can be competition to *HEA*.

b. There are other parameters such as polydispersity ($\delta$), valence electron concentration (VEC), the chemical enthalpy of mixing ($\Delta H_{mix}$) and *HEA* formability parameter ($\Omega$) which have been proposed for phase selection between amorphous and HEA [12, 13, 17]. However, unlike *PSV*, these parameters cannot predict the *A-HEA* transformation as a function of temperature. In equation 7, the *PSV* is dependent on two terms – elastic mismatch enthalpy term and the topological enthalpy term. The elastic mismatch enthalpy term is related to elastic moduli of the component elements which in turn is known to change with temperature. The topological enthalpy term on the other hand is independent of temperature. Therefore, using the *PSV*, the possibility of *A-HEA* transformation can be checked for temperatures greater than $T_g$.

While the *A-HEA* transformation can be understood through Miedema's model, the phase selection between *HEA* and intermetallics/compounds for those systems which have *PSV* negative can be determined using the perturbation model proposed by Luan et al. [18]. Unlike the Miedema's model, the perturbation model allows for quick determination of the phase selection between *HEA* and intermetallics/compounds. Further details of the perturbation model are provided in the supplementary section.

With the objective of identifying multi component systems which are amorphous at ambient temperatures but capable of converting into *HEA* above $T_g$, we start by investigating popular *BMGs*. The rationale for choosing *BMGs* for this analysis is that these can be produced in bulk form and they have wider commercial applications compared to other amorphous alloys. With this in mind, the *PSV* value of the *BMGs*, serial # 1 to 5, reported in Table 1 was determined at both room temperature as well as at temperature equal to $T_g + 20$ Kelvins. The $T_g$ values for these *BMGs* were taken from the literature and are reported in Table 1 [19-23]. As shown in Table 1, the reported *BMGs* have a positive *PSV* both at room temperature and at $T_g+20$ K indicating that for these systems *HEAs* are thermodynamically unstable compared to the parent amorphous phase. Hence these *BMGs* are incapable of crystallizing into a *HEA* at temperatures above $T_g$. In Table 1, are also reported seven compositions which we propose are capable of undergoing the *A-HEA* transformation at $T_g+20$ K.

Through our calculations we found that systems capable of *A-HEA* transformation should have *PSV* value in the range of 0-0.5 kJ/mol at room temperature. Since the *PSV* value is strongly influenced by the elastic enthalpy which in turn is dependent on the difference in molar volume of the constituent elements the distribution of the molar volume of the constituent elements is crucial. We introduce a parameter $\Gamma$ called the molar volume dispersity parameter to describe the molar volume distribution. This parameter is similar to that of the polydispersity parameter $\delta$ proposed by Guo et al. [12]. The parameter $\Gamma$ can be determined using $\Gamma = \sqrt{x_i \cdot \left(1 - \frac{V_{m,i}}{\overline{V_m}}\right)^2}$, where $\overline{V_m}$ is given by $\sum x_i V_{m,i}$ where $V_{m,i}$ is the molar volume of the *i*th component and $\overline{V_m}$ is the average molar volume of the system and $x_i$ is the atomic concentration of the *i*th element. We believe that compositions capable of the *A-HEA* transformation would bear a $\Gamma$ in the range of 0.2 to 0.3, as shown in Table 1. However, the $\Gamma$ is a good as a screening parameter only and has to be used in conjunction with the *PSV* for identifying systems capable of *A-HEA* transformation.

**Table 1:** *PSV* and *Γ* for BMGs and our predicted system

| Sl. # | Alloy Systems | *PSV* at RT kJ/mol | *PSV* at ($T_g$+20) K kJ/mol | Γ | Ref |
|---|---|---|---|---|---|
| 1 | $Zr_{41.2}Ti_{12.8}Cu_{12.5}Ni_{10}Be_{22.5}$ | 15.76 | 15.23 | 0.39 | [19] |
| 2 | $Sr_{20}Ca_{20}Yb_{20}Mg_{20}Zn_{20}$ | 7.88 | 7.46 | 0.41 | [20] |
| 3 | $La_{55}Al_{25}Co_5Cu_{10}Ni_5$ | 13.21 | 12.94 | 0.45 | [21] |
| 4 | $Nd_{60}Al_{15}Ni_{10}Cu_{10}Fe_5$ | 11.40 | 11.17 | 0.39 | [22] |
| 5 | $Ni_{53}Nb_{20}Ti_{10}Zr_8Co_6Cu_3$ | 5.82 | 5.39 | 0.28 | [23] |
| 6 | $Zr_{20}Mo_{15}Nb_{16}Mn_{30}Cr_{19}$ | 0.26 | -0.21 | 0.27 | |
| 7 | $Zr_{35}Mn_{30}V_{16}Ti_9Cr_{10}$ | 0.40 | -0.16 | 0.29 | |
| 8 | $Ti_{40}Fe_{27}Co_{17}W_6Be_{10}$ | 0.31 | -0.12 | 0.24 | |
| 9 | $Ti_{40}Fe_{23}Zn_{10}Mg_{20}Be_7$ | 0.30 | -0.10 | 0.27 | |
| 10 | $Co_{25}Ni_{25}Fe_{19}B_{23.5}Si_{7.5}$ | 0.26 | -0.10 | 0.28 | |
| 11 | $Ni_{30}Fe_{16.5}Mn_{25.5}B_{14.5}Si_{13.5}$ | 0.20 | -0.16 | 0.29 | |
| 12 | $Ni_{31}W_{16.5}Mn_{30}B_{14.5}Si_8$ | 0.25 | -0.23 | 0.27 | |

Blue colored multi-component systems are known BMG systems reported in literature and green colored systems are our predicted system that are capable of crystallizing into HEA from amorphous phase.

It is interesting to note that most of the *BMGs* reported in Table 1 have *Γ* greater than 0.3. It is also apparent from Table 1 that systems with *Γ* > 0.3 have a high value of *PSV* which means they are very less likely to undergo *A-HEA* transformation. This applies very well to *Vit 1* which is known to crystallize into intermetallics / compounds instead of *HEAs* during annealing at temperatures above $T_g$. For the predicted systems (serial # 6 to 12 in Table 1), the $T_g$ value is not known *apriori*. Lu and Li [24] have suggested that for an amorphous system, $T_g = 0.385 T_m$, where $T_m$ is the melting point of amorphous system guided by the *rule of mixtures*. To validate this correlation and with an objective of determining the $T_g$ of the systems # 6 to # 12, we calculated the $T_m$ for a number of amorphous systems (whose experimentally determined $T_g$ is reported in literature and shown in supplementary Table S1) and plotted it against the $T_g$ calculated using the Lu and Li scheme. Figure 1 shows a plot of calculated and experimental $T_g$ of a variety of amorphous systems against the system $T_m$. This plot shows that there is a good agreement between actual and calculated $T_g$ and by and large the $T_g$ of an amorphous system is found to fall in the range of 0.35 – 0.4 $T_m$. Using this knowledge, we estimated the $T_g$ for the predicted systems and determined their *PSV* at $T_g+20$ K. These systems supported the *A-HEA* transformation $T_g+20$ K.

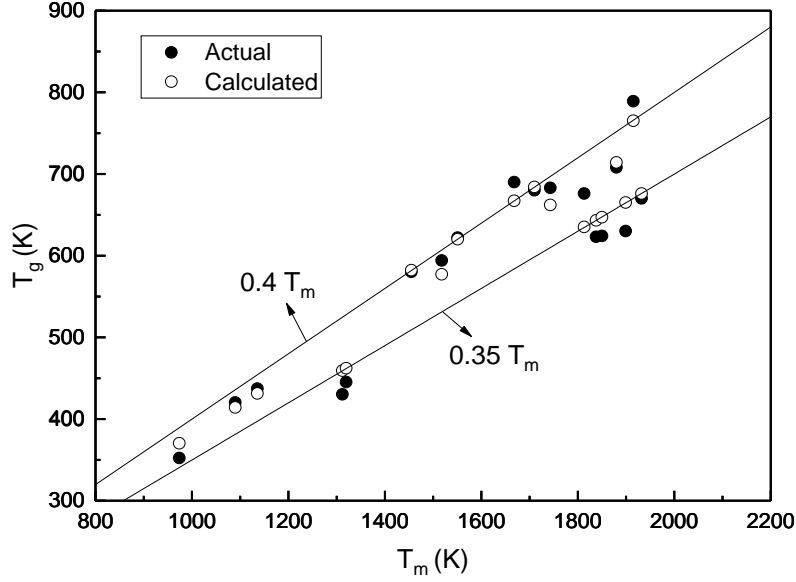

**Figure 1:** Comparison of calculated and actual Tg for a variety of amorphous systems. The compositions of these systems and their actual $T_g$ is given in Table S1 [19, 20, 25-36].

Since these compositions support *A-HEA* transformation, we were keen to understand their location in the plot of $\Delta H_{mix}$ and $\delta$ proposed by Guo et al. [12]. In addition to alloy systems mentioned in Table 1, few popular *HEAs* (compositions shown in supplementary Table S2) were also placed in Guo et al. plot to highlight contrast in their location compared to *BMGs*. As per Guo et al. [12], top left quadrant is assigned to *HEA* forming compositions and the bottom right quadrant is assigned to *BMG* forming compositions. Indeed, all BMGs (serial # 1 to 5), except one, reported in Table 1 are located in the bottom right quadrant. As shown in Figure 2, it is observed that *A-HEA* supporting compositions are scattered within the bottom and top right quadrants suggesting that Guo et al. [12] plot while guiding phase selection between Amorphous and *HEA* forming compositions, may not be able to predict the *A-HEA* transformation. Instead based on our observations, we believe that the *PSV* is a more appropriate parameter for predicting the *A-HEA* transformation. In support of our assertion, we plot the *PSV* value of *Vit 1* and the predicted composition (serial # 7 $Zr_{35}Mn_{30}V_{16}Ti_9Cr_{10}$) at room temperature (RT) and at $T_g + 20$ K in Figure 3. The *PSV* plots of all compositions reported in Table 1 are shown in supplementary Figure S1. As Figure 3 reveals, the strong glass formers such as *BMGs* bear a high positive value of *PSV* at RT and undergo a small decrease of *PSV* at $T_g$ +20 K implying that they will remain the stable phase vis-à-vis *HEA*. On the other hand, the predicted systems have a small positive value of *PSV* at RT which changes to a negative value at $T_g$+20 K implying the capability of the *A-HEA* transformation.

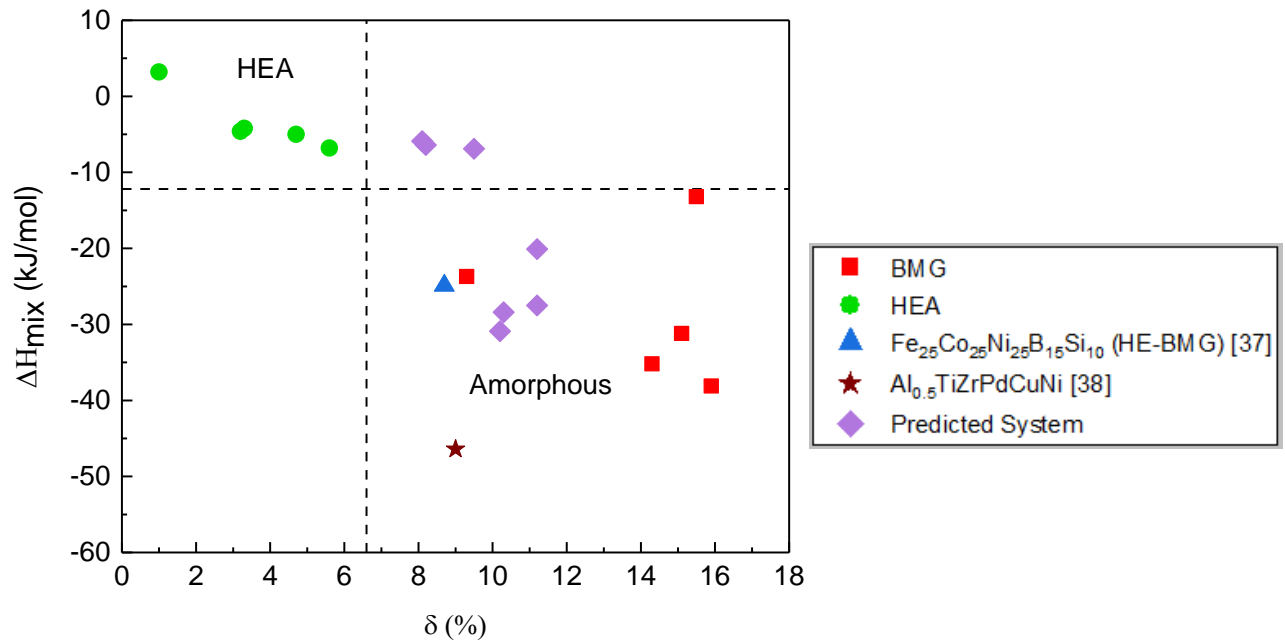

**Figure 2:** Plot between $\Delta H_{mix}$ vs. $\delta(\%)$ for multi-component systems discussed in the Table 1.

It is apparent that systems capable of *A-HEA* transformation should have a small positive value of *PSV* at *RT* ensuring that the *A-HEA* transformation is not very difficult.

As a further validation of the *PSV* approach, we plot *PSV* value of a particular High Entropy Bulk Metallic Glasses (*HE-BMG*) bearing a composition $Fe_{25}Co_{25}Ni_{25}B_{15}Si_{10}$ at *RT* and $T_g$+20 K in Figure 3 [37]. This composition according to Guo et al. plot [12] (shown in Figure 2) should exist as an amorphous phase and it is indeed so when it is cast at high cooling rates. However, the same composition prefers forming a *HEA* phase (largely a mixture of *fcc* phase with an unknown compound) instead of an amorphous phase when cooling rates are lower [37]. This seems to be in good agreement with the predictions of the *PSV* approach which suggests formation of a *HEA* based on the negative value of *PSV* at *RT*. This implies that this composition is supposed to exist as a *HEA* at room temperature under equilibrium conditions as correctly captured by *PSV* approach.

As a further validation, we calculated the value of *PSV* for Al$_{0.5}$TiZrPdCuNi system suggested by Takeuchi et al. [38] and found it to be -1 kJ/mol at *RT*. It is interesting to note that $\Delta H_{mix}$ and $\delta$ for Al$_{0.5}$TiZrPdCuNi is -46.7 kJ/mol and 8.8 % respectively and this combination of values makes it fall within the amorphous domain suggested by Guo et al. [12]. However, this particular system formed an amorphous phase only when processed by melt spinning but formed a *bcc* solid-solution when processed through copper-die casting. This is in contrast to the predictions of Guo et al. [12] who suggested that $\Delta H_{mix}$ should not be more negative than -25 kJ/mol for solid-solution formation [12]. However, *PSV* approach captures correctly that the Al$_{0.5}$TiZrPdCuNi system [38] will prefer *HEA* over amorphous phase under low cooling rates. This, indicates that the *PSV* in addition to identifying *A-HEA* compositions can help predicting the preferred phase.

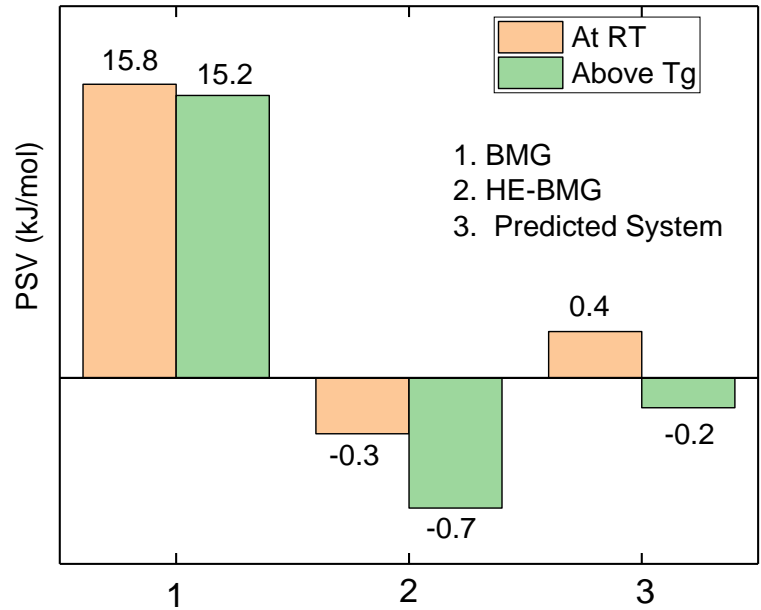

**Figure 3:** Plot for *PSV* value at room temperature and temperature above *T$_g$* for a popular BMG (Vitreloy-1) [19], HE-BMG (Fe$_{25}$Co$_{25}$Ni$_{25}$B$_{15}$Si$_{10}$) [37] and one of our predicted system (Zr$_{35}$Mn$_{30}$V$_{16}$Ti$_9$Cr$_{10}$)

In summary, we propose a model to identify multi component compositions capable of undergoing an amorphous to HEA transformation. The thermodynamics based model guided by both the Miedema's model and perturbation model was able to identify seven amorphous compositions which prefer to crystallize into *HEAs* instead of intermetallics / compounds. Compositions bearing a *PSV* value in the range of 0 to 0.5 kJ/mol and a $\Gamma$ value of 0.2-0.3 appear to encourage the *A-HEA* transformation and also help in predicting phase stability.